\documentclass[prl,amsmath,amssymb,twocolumn,superscriptaddress]{revtex4}

\usepackage{amsmath,amssymb}
\usepackage[usenames]{color}
\usepackage{mathbbol}
\usepackage{amssymb}
\usepackage{grffile}
\usepackage[pdftex]{graphicx}
\usepackage{amsmath, amstext, amssymb, amsfonts, amsxtra}
\usepackage{textcomp}
\usepackage{xspace}
\DeclareSymbolFontAlphabet{\amsmathbb}{AMSb}

\newcommand{\aop}{\hat{a}} 
\newcommand{\bop}{\hat{b}} 
\newcommand{\cop}{\hat{c}}
\newcommand{\adop}{\hat{a}^{\dagger}} 
\newcommand{\bdop}{\hat{b}^{\dagger}} 
\newcommand{\cdop}{\hat{c}^{\dagger}}
\newcommand{\afop}{\hat{\alpha}} 
\newcommand{\afdop}{\hat{\alpha}^{\dagger}} 
\newcommand{\rhoop}{\hat{\rho}}

\newcommand{\Mm}{\bold{M}}
\newcommand{\Um}{\bold{U}}
\newcommand{\Tm}{\bold{T}}

\newcommand{\Hop}{\hat{H}}
\newcommand{\spec}{\mathcal{J}}

\newcommand{\im}{{\rm i}}
\newcommand{\hc}{{\rm H.c.}} 
\newcommand{\diag}{{\rm diag}}

\newcommand{\wm}{\omega_{\rm max}} 
\newcommand{\Nm}{N_{\rm max}} 
\newcommand{\Js}{\mathcal{J}} 
\newcommand{\Hdtp}{\Hop^{{\rm dis},\prime}_{\rm tot}}

\newcommand{\sutd}{EPD Pillar, Singapore University of Technology and Design, 8 Somapah Road, 487372 Singapore} 
\newcommand{\munich}{Department of Physics and Arnold Sommerfeld Center for Theoretical Physics, Ludwig-Maximilians-University Munich, 80333 Munich, Germany}

\begin{document}

\title{Stable-unstable transition for a Bose-Hubbard chain coupled to an environment}                     

\author{Chu Guo} 
\affiliation{\sutd}
\author{Ines de Vega}  
\affiliation{\munich} 
\author{Ulrich Schollw\"ock}  
\affiliation{\munich} 
\author{Dario Poletti}
\affiliation{\sutd}

\begin{abstract} 
Interactions in quantum systems may induce transitions to exotic correlated phases of matter which can be vulnerable to coupling to an environment. Here, we study the stability of a Bose-Hubbard chain coupled to a bosonic bath at zero and non-zero temperature. We show that only above a critical interaction the chain loses bosons and its properties are significantly affected. The transition is of a different nature than the superfluid-Mott insulator transition and occurs at a different critical interaction. 
We explain such a stable-unstable transition by the opening of a charge gap. The comparison of accurate matrix product state simulations to approximative approaches that miss this transition reveals its many-body origin.   
\end{abstract}

\pacs{05.30.Jp, 03.65.Yz, 67.10.Jn, 64.70.Tg}

\maketitle

 
Interactions lead to emergent collective phenomena and quantum phase transitions in soft and condensed matter systems. However, real  systems are often coupled to an environment, and therefore we need to understand how this affects the system properties. The environment may destroy key properties of the isolated system e.g.\ due to decoherence or to particle losses. Recent experiments with ultracold atoms have started investigating these phenomena \cite{Ott1, Vengalattore, Ott2, Schneider, Takahashi}. The interplay between system and environment can even be tailored to generate interesting non-equilibrium phases of matter \cite{Diehl2008}.     

In this work, we focus on the robustness of a manybody quantum system against the coupling to an external bath: We consider a 
one-dimensional Bose-Hubbard chain (BHC) which exhibits a quantum phase transition between a superfluid phase and a Mott-insulating phase driven by interaction strength and particle density \cite{Giamarchi1987, FisherFisher1989, jaksch1998}. 
We study the dissipative dynamics of the BHC after coupling its last site to a bosonic bath either at zero or non-zero temperature. 
Dissipation is often modeled by a Lindblad master equation \cite{lindblad1976, gorini1976} which implies a number of assumptions on the system, the bath(s) and their coupling \cite{breuerbook, quantumnoise, weissbook}. The Lindblad master equation assumes weak coupling to and instantaneous recovery of the environment (Markov approximation), which makes the system-environment state separable (Born approximation). If separability is lost, one has to go beyond a Lindblad master equation approach. 
For this reason, the Hamiltonian dynamics of the system plus bath has been studied recently in spin systems \cite{mascarenhas2017, mascarenhas2017b}. In other cases, the dynamics of two connected spin or bosonic chains prepared in different states was investigated \cite{Ponomarev2011, DeLucaRossini2014, BiellaFazio2016, Ljubotina2017}. 

Convenient methods to study both equilibrium and non-equilibrium properties of 1D systems are matrix product states (MPS) based algorithms which can be modified to include the dissipative effects of an environment as described by the Lindblad formalism or the equivalent Markovian quantum jump approach \cite{scholl2011,verstraete2004,zwolak2004,Daley2014}.
Here we use an MPS algorithm to study an interacting bosonic system coupled to a bosonic bath at either zero or finite temperature. Our analysis is, to the best of our knowledge, the first to tackle this problem without assuming a Born-Markov approximation. 
To facilitate the numerics we characterize the bath thermal state by performing a thermofield transformation \cite{devegabanuls2015}. Moreover, we consider a unitary transformation that maps the environment into a chain structure containing only nearest-neighbor tunnelling \cite{Wilson1975, prior2010,chin2010b}. 

We show that for a zero-temperature bath the system is stable against the dissipation only if the interaction is below a certain threshold, while the system loses bosons for a larger interaction. For non-zero temperatures the evolution changes drastically depending, again, on the strength of the interaction. In this case the system's number of bosons can increase on numerically accessible time scales for sufficiently weak interactions. These effects are purely many-body and beyond weak-coupling: we show that both a weak-coupling and a mean-field approach give qualitatively different results. We should also stress that this stable-unstable transition produced by the system-bath coupling, is of different nature than the ground-state superfluid to Mott-insulator phase transition, and in fact occurs at a different critical interaction strength. 

{\it The model}: 
We consider a BHC coupled to an environment of free bosonic oscillators. The Hamiltonian of the system plus environment can be written as
\begin{align}
\Hop_{\rm tot} = \Hop_S + \Hop_E + \Hop_I,
\end{align}
with $\Hop_S$ being the Hamiltonian of a BHC of length $L$, tunnelling amplitude $J$ and onsite interaction strength $U$
\begin{align}
\Hop_S =& - J \sum_{j=1}^{L-1}(\afdop_{j} \afop^{}_{j+1} + \hc) + \frac{U}{2}\sum_{j=1}^L \afdop_j\afop^{}_j(\afdop_j\afop^{}_j - 1). 
\end{align} 
To have lighter notations, henceforth we work in units such that $J=\hbar=k_B=1$. 
We are interested in the relaxation dynamics of the system starting from the ground state of average filling $\bar{n} = 1$, which we denote as $\vert E^0_{N=L} \rangle$, where $N$ is  the total number of bosons in the system. In $1D$ the transition from superfluid to Mott-insulator occurs at the critical value $U_c \approx 3.37$ \cite{KuhnerMonien2000}. 
The $L$-th site of the BHC is coupled to an environment of harmonic oscillators, whose Hamiltonian can simply be written as $\Hop_E = \int\! d\omega \; \omega \bdop_{\omega} \bop^{}_{\omega}$. We consider a coupling between the system and the environment of the form
\begin{align}
\Hop_I = \int \! d\omega  \sqrt{\Js(\omega)} \left(\afdop_L \bop^{}_{\omega} + \afop^{}_L \bdop_{\omega} \right) \label{eq:sb_coupling}       
\end{align}
where $\Js(\omega) = g \omega^{\eta}$ is the spectral density and it corresponds to a sub-ohmic, ohmic, or super-ohmic bath respectively for $\eta <1$, $\eta = 1$ or $\eta >1$ \cite{weissbook, devega2015c}. We choose a sharp cut-off of the spectral density $\wm$ such that $\Js(\omega)=0$ for $\omega>\wm$. The environment is prepared in a thermal state with temperature $T$, i.e. $\rhoop_E \propto e^{- \Hop_E/ T}$.

{\it Method}: To apply MPS methods we need to discretize the environment. We use a linear discretization of the bath \cite{devega2015b} into $\Nm$ oscillators evenly spaced by $\Delta\omega = \wm/\Nm$,  resulting in
$\Hop_E^{\rm dis} = \sum_{j=1}^{N_{\rm max}}\omega_j \bdop_j \bop^{}_j$, with $\omega_j = j\Delta\omega$. At the same time, $\Hop_I$ becomes $\Hop_I^{\rm dis} = \sum_{j=1}^{\Nm}\sqrt{\spec_j}(\afdop_L\bop^{}_j + \afop^{}_L\bdop_j)$, with $\spec_j = \int_{\omega_j}^{\omega_{j+1}} d\omega \spec(\omega) \approx \spec(\omega_j) \Delta\omega $ and we have used $\bop_j\equiv\bop_{\omega_j}$. We have tested the convergence for different values of $\Delta \omega$ choosing, henceforth, $\Delta \omega=0.01$ and $\wm=6$.

For a bath at temperature $T=0$, we map the bath to a long linear non-interacting bosonic chain with nearest-neighbor couplings \cite{prior2010, chin2010b, devega2015b, supp}.  The full Hamiltonian of the system is therefore represented as a long chain in which the first $L=10$ sites correspond to the BHC and the next sites ($200$ in our simulations) correspond to the transformed environment oscillators. The full system wave function is then evolved accordingly from the initial condition $\vert\psi\rangle = \vert E^0_{N=L} \rangle \otimes \vert 0\rangle_c$ where $\vert 0 \rangle_c$ is the vacuum of the $\cop_j$, the annihilation operators of free bosons on the chain representing the discretized bath \cite{supp, num_cons}. We use a second-order Suzuki-Trotter split-step method with time step $dt=0.01$, bond dimension $4000$, and local basis dimension $5$ in the bath; in the system it is $7$ for $U=1$, $6$ for $ 1<U \leq 4$, $5$ for $U>4$. 

For $T>0$, we first perform a thermofield transformation \cite{devegabanuls2015}, in which the finite temperature environment is exactly mapped to two virtual environments at zero temperature. These two environments are then unitarily transformed to two different chains of oscillators having nearest-neighbor coupling and annihilation (creation) operators $\aop_{1,j}$ and $\aop_{2,j}$ ($\adop_{1,j}$ and $\adop_{2,j}$) respectively. 
The total state to be evolved can then be written as $\vert\psi\rangle = \vert E^0_{N=L}\rangle \otimes \vert 0 \rangle_{a_1} \otimes \vert 0\rangle_{a_2}$, where $\vert 0 \rangle_{a_1}$, $\vert 0 \rangle_{a_2}$ are the vacuum states of all the $\aop_{1, j}$ and $\aop_{2, j}$ corresponding to the two thermofield environments (see \cite{devegabanuls2015, supp} for details). The parameters used for the simulations are the same as for $T=0$ except that we use a non-number conserving algorithm with a bond dimension $300$, and swap gates \cite{SteveWhite2010} to implement the $2$nd order Suzuki-Trotter evolution.

%
\begin{figure}
\includegraphics[width=\columnwidth]{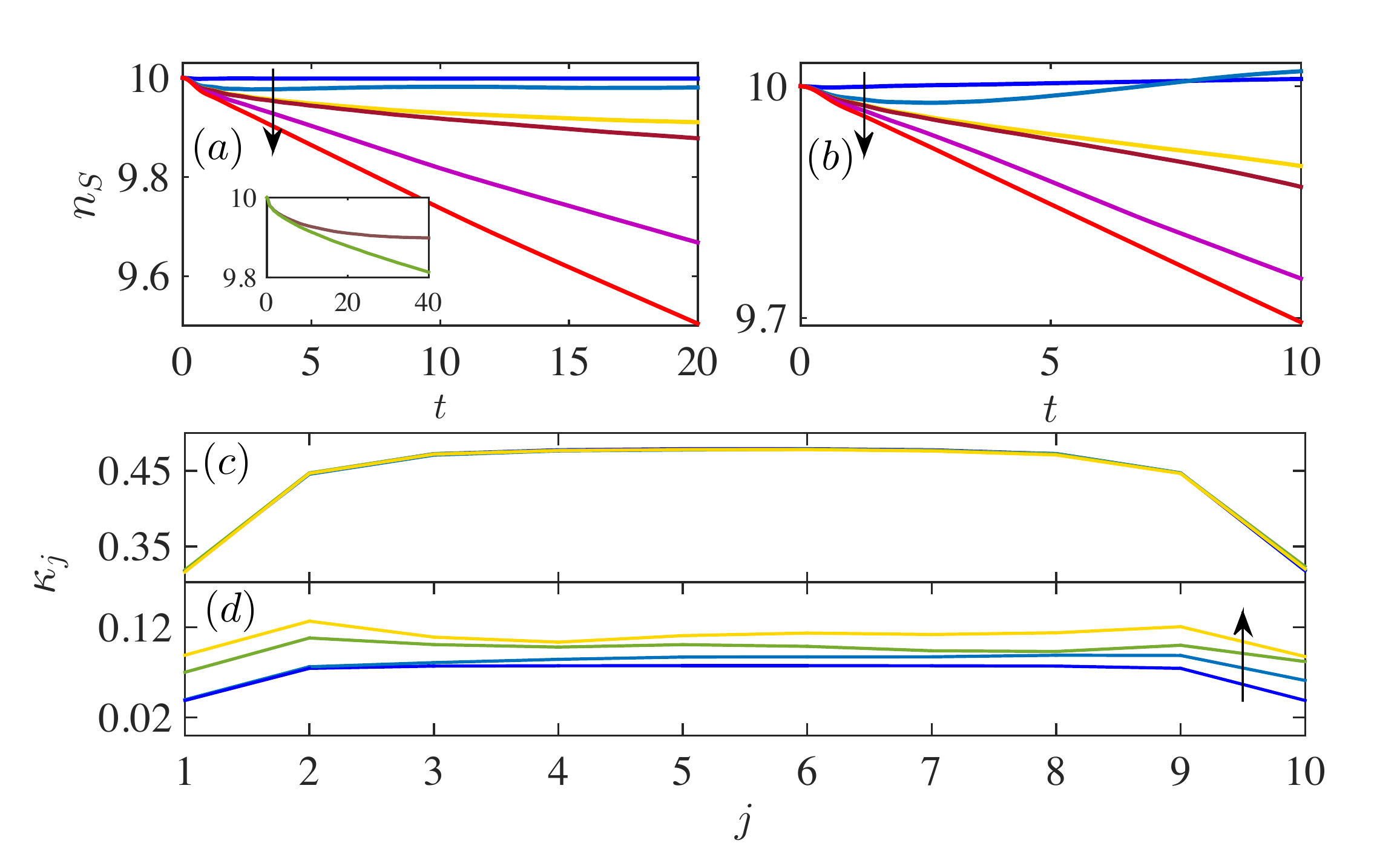}
\caption{(color online) (a,b) Total number of bosons in the system as a function of time $t$ for (a) $T=0$ and (b) $ T=0.1$. The lines in the direction of the arrow correspond to $U=0,\; 2,\; 2.8,\; 2.9,\; 4,\; 10$. The inset of (a) shows a detail for $U=2.8$ and $2.9$ for longer times. (c,d) Local fluctuations $\kappa_j$ as a function of site $j$, for $U=2$ and $U=10$, for different times: in the direction of the arrow $t=0, 4, 10, 16, 20$ (in (c) the lines are on top of each other). In all panels, the other bath parameters used in both cases are $g=0.01$  and $\eta=0.5$.} \label{fig:fig1} 
\end{figure}

{\it Stability analysis of the manybody system}: We analyze the stability of the ground state of the system by monitoring the number of bosons within the BHC 
\begin{align}
n_S(t) = \sum_{j=1}^L \langle \psi(t) \vert \afdop_j \afop^{}_j \vert \psi(t) \rangle.  
\end{align}

As shown in Fig.\ref{fig:fig1}(a,b), when the interaction strength $U$ varies the evolution of $n_S(t)$ changes substantially. To understand this, we notice that the system Hamiltonian, $\Hop_S$, conserves the total number of bosons. The environment couples different number sectors and thus breaks the number conservation of the system. Since initially $N=L$, and given the coupling (\ref{eq:sb_coupling}), the environment will first induce transitions to states with $N=L\pm 1$ bosons. 

At $T=0$, see Fig.\ref{fig:fig1}(a), the environment does not have bosons to transfer to the system 
so that it will only be able to couple the initial state to states with $N=L-1$ bosons. The key is thus to study the energy difference between the ground state energy of the system $E^0_{N=L}$, and the ground state energy $E^0_{N=L-1}$ corresponding to $N=L-1$ atoms, i.e. $\Delta E= E^0_{N=L} - E^0_{N=L-1}$. This is the largest amount of energy that the system can lose when the first boson is removed.     
For large values of $U$, $\Delta E>0$. The transfer of a boson from the system to the bath can occur because the system loses energy while the bath gains energy and hence the overall energy of the system plus bath is conserved. 
However at low interaction strength $U$, $\Delta E<0$ and hence the dynamics of the overall Hamiltonian system is almost completely frozen \cite{SmallMod}.  

The different response of the system to the bath is also well evidenced by the system's local fluctuations $\kappa_j=\langle (\afdop_j\afop^{}_j)^2\rangle - \langle \afdop_j\afop^{}_j \rangle^2 $. For low interaction, Fig.\ref{fig:fig1}(c) with $U=2$, the fluctuations change minimally (the curves for different times are superimposed), while for larger interactions, Fig.\ref{fig:fig1}(d) with $U=10$, there is a sizable ``fluctuation wave'' starting from site $j=10$, where the bath is connected, and propagating through the system.    

\begin{figure}
\includegraphics[width=\columnwidth]{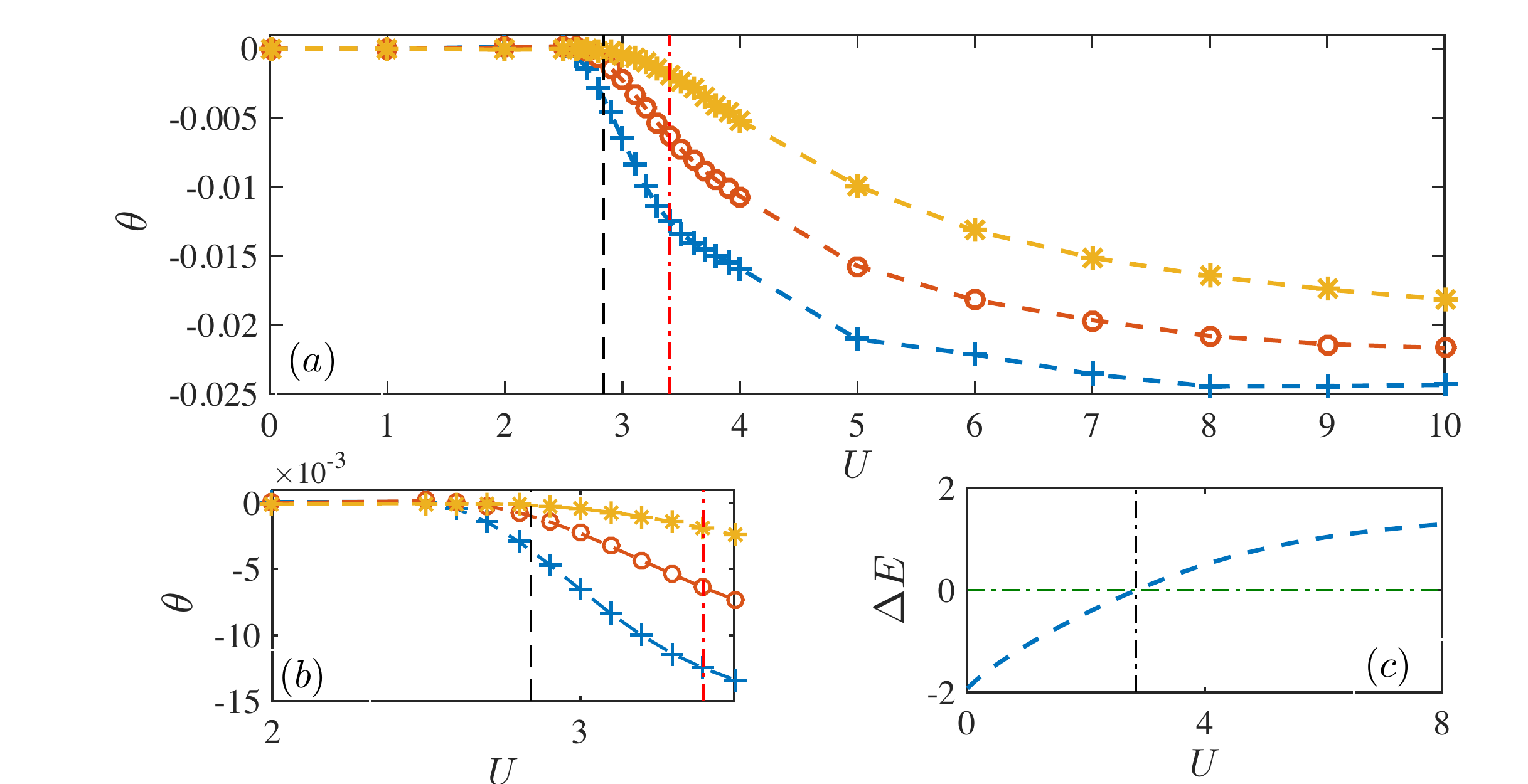}
\caption{(color online) (a) $\theta$, the slope of the evolution of $n_S(t)$ as a function of $U$. Each line corresponds to a different spectral density: blue crosses for $\eta=0.5$ (sub-ohmic), red circles for $\eta=1$ (ohmic) and yellow stars for $\eta=2$ (super-ohmic) while $g=0.01$, $0.01$, and $0.01$ respectively. The vertical black-dashed line highlights the critical value of the interaction $U=U_s$ while the red dot-dashed line shows $U_c$, the location of the quantum phase transition between superfluid and Mott-insulator. (b) Detail of (a) near the stable/unstable critical interaction. (c) Groundstates energy difference $\Delta E$ versus $U$.} \label{fig:fig2} 
\end{figure}

The transition from stable to unstable dynamics is clearly highlighted in Fig.\ref{fig:fig2}(a,b) where we plotted the decay slope $\theta$ from a linear fit of $n_S$ for $2<t<20$ for various values of the interaction strength $U$. Each line corresponds to a different type of bath, sub-ohmic ($\eta=1/2$), ohmic ($\eta=1$) and super-ohmic ($\eta=2$). 
In all scenarios there is a clear transition between non-decaying and decaying dynamics. Computing $\Delta E=0$ for systems up to $L=120$, we identify the transition to occur at $U_s\approx 2.82$ \cite{evo_ground}. This transition line is highlighted by a black-dashed line in Fig.\ref{fig:fig2}. With the red-dotted line instead we show $U_c$, the critical interaction strength for the superfluid-Mott insulator transition, which is clearly larger than the interaction strength at which the stable-unstable transition occurs. In Fig.\ref{fig:fig2}(c), by plotting $\Delta E$ vs $U$ for $L=10$, we clearly show that $\Delta E>0$ for large $U$ and it is negative for smaller interaction strength \cite{N10}.        
In Fig.\ref{fig:fig2}(b) we zoom into Fig.\ref{fig:fig2}(a) around the transition point. 
Since the coupling to the environment produces a shift in the energy levels, the transition point may be shifted. Such shift is particularly significant for the sub-ohmic bath, as it effectively produces a stronger coupling. 

The dynamics ensuing the coupling to a $T>0$ bath presents important similarities and differences compared to a $T=0$ bath. For large interactions the number of bosons still decreases in a similar manner. For small interactions it is still not possible for the system to lose bosons. However, since the bosonic bath is populated, there are energy conserving processes for which the number of bosons in the system increases, as shown in Fig.\ref{fig:fig1}(b), while the bath loses bosons.

%

\begin{figure}
\includegraphics[width=\columnwidth]{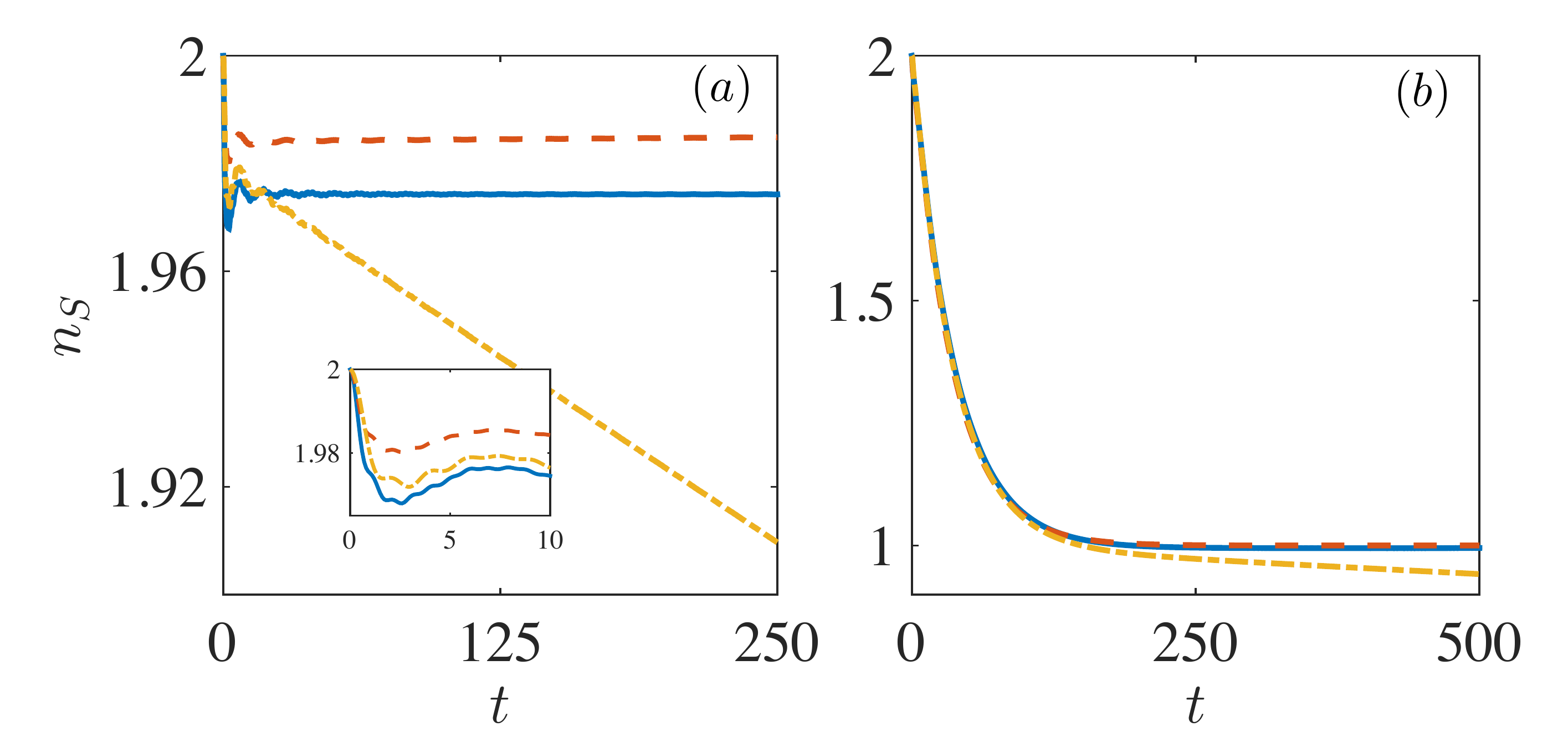}
\caption{(color online) Total number of bosons in the system as a function of $t$ for (a) weak interaction $U=1$ (the inset focuses on the short-time dynamics) and (b) strong interaction $U=5$. The environment has temperature $T=0$, and we have used $g=0.01$ and $\eta=0.5$. The blue continuous line, dashed red line, and yellow dot-dashed line correspond respectively to the MPS results, the effective model, and the Redfield equation.        
} \label{fig:fig3} 
\end{figure}

{\it Long time dynamics and comparison to weak-coupling master equation}: 
For smaller systems we can study the dynamics for much longer times, e.g. up to $t=500$. The case for $T=0$ and $N=L=2$ is plotted with a blue continuous line in Fig.\ref{fig:fig3} for small (a) and large (b) interactions. 

For such a system we can readily study the time evolution of the reduced density matrix of the system as given by a second order weak coupling master equation (hereby denoted as Redfield equation) which requires the Born but not the Markov approximation.
As shown in Fig.\ref{fig:fig3} (see also the inset), the Redfield equation is accurate at short times but predicts the wrong steady state both for weak and strong interactions $U$, even though the coupling between the system and the environment is relatively weak, $g=0.01$. The disagreement is particularly important for weak interactions for which the exact dynamics predicts a stable dynamics while the Redfield master equation predicts a decay. For stronger interactions, for which the system can lose bosons, the Redfield equation is accurate up to much longer times. 

\begin{figure}
\includegraphics[width=\columnwidth]{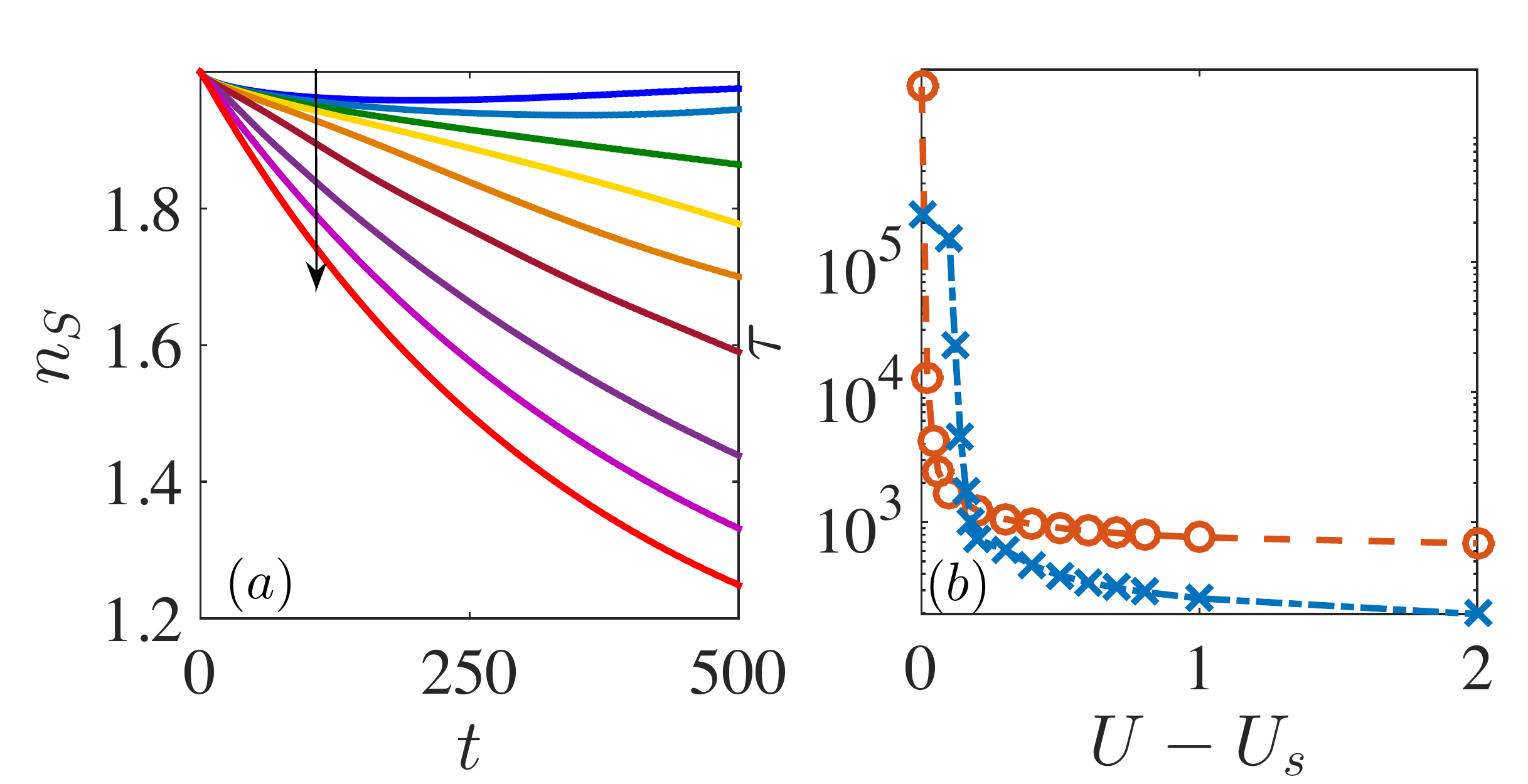}
\caption{(color online) (a) Total number of bosons in the system as a function of $t$ for $T=0, \; g=0.001$. The lines in the direction of the arrow correspond to $U=3, 3.02, 3.04, 3.06, 3.1, 3.2, 3.5, 4, 5$. (b) The relaxation time $\tau$ as a function of $U-U_s$. The dot-dashed line with blue crosses corresponds to $g=0.01$, while the dashed line with red circle corresponds to $g=0.001$.
} \label{fig:fig4} 
\end{figure}

In order to obtain qualitatively correct results, an approach which correctly accounts for the system-bath correlations that are built up is needed. For the two-site case, where only a few energy levels are relevant for the dynamics, we can use an effective model based on the ground states corresponding to $N=1$ and $N=2$ bosons in the system (see \cite{supp}). The prediction of this effective model is shown with red dashed lines in Fig. \ref{fig:fig3} and indeed matches qualitatively the exact numerical results of MPS simulations. 
The coupling strength $g$, and the particular typology of the spectral density $\mathcal{J}$, do not have a qualitative effect on the stable-unstable transition, which only becomes sharper for weaker coupling. In Fig.\ref{fig:fig4}(a) we show the number of bosons for different interactions strengths. We then fit these curves with an exponential decay $n_S(t) -n_S({\infty}) \propto e^{-t/\tau}$ with $n_S({\infty})=1$ to estimate the time scale $\tau$. In Fig.\ref{fig:fig4}(b) we plot this time scale versus the interaction $U$ minus the critical interaction $U_s$, both for $g=0.01$ ($\times$ on blue dot-dashed line) and for $g=0.001$ (circles on red dashed line). Clearly the time scale changes dramatically as the interaction approaches $U_s$, especially for $g=0.001$, indicating that a very different dynamics occurs for lower interactions.

It is important to stress that this stable-unstable transition cannot be predicted by simpler meanfield approaches. For instance, a Gutzwiller ansatz \cite{RoksharKotliar1991, krauth1992, jaksch1998} to study a Mott-insulator coupled to a $T=0$ bath predicts a completely frozen dynamics with no decay, which is qualitatively inaccurate \cite{meanfield}.

%

{\it Conclusions}: To summarize, we have studied the stability of a Bose-Hubbard chain when coupled to a thermal bosonic bath. We have shown that at $T=0$, when varying the on-site interaction strength between the bosons in the BHC across a critical value $U_s$, there is a transition between a stable and an unstable dynamics. This transition is due to a change in sign of the difference between the ground state energies of a system with $N=L$ (corresponding to the initial state) and $N=L-1$ bosons. 
The stable-unstable transition occurs at lower values of the interaction compared to the equilibrium quantum-phase transition between a superfluid and a Mott-insulator, showing that the two transitions, while both due to the on-site interaction $U$, are of distinct nature.  
We have also shown that these effects go beyond a weak-coupling master equation description, and also a simple meanfield approach, because both methods predict qualitatively incorrect results. We can therefore conclude that this is a many-body and strong-coupling effect.  

This system can be realized experimentally with atoms in two hyperfine states, one which is trapped by a lattice and one which is not and forms a reservoir as described in \cite{devega2008}. In particular the reservoir atoms would be confined to the end of the lattice as for example in \cite{krinner2017}, although with bosons. Alternatively, the bosonic bath could be excitations within a BEC while the system is formed by impurities \cite{recati2005}. It would also be possible to realize experimentally the description of our system after the star-to-chain transformation. 
In this case the bosons would be the local vibronic excitations of the ions which can be made locally interacting thanks to transverse laser beams \cite{porrascirac2004, duttasengupta2013}. The system would be the part of the ion chain with non-zero interactions while the non-interacting portion, with suitable tunnelling couplings, would emulate the bath.\\

The stable-unstable transition may also signal the presence of a non-equilibrium phase transition in the steady state. In fact the properties of the steady state will be markedly different in the two sides of the transition, as it is highligted for instance by the different total number of bosons in the system. However our current tools do not allow us to reach the steady state for a large number of atoms and therefore show this. Future work may consider systems with richer phase diagrams and more exotic phases compared to a Bose-Hubbard chain.

{\it Acknowledgments}: D.P. acknowledges support from Ministry of Education of Singapore AcRF MOE Tier-II (project MOE2016-T2-1-065, WBS R-144-000-350-112) and fruitful discussions with B. Gr\'emaud. D.P. was hosted by ICTP during part of this study. This work was performed in part at the Aspen Center for Physics, which is supported by NSF grant PHY-1607611. I. de V. further acknowledges support by the DFG-grant GZ: VE 993/1-1.

\setcounter{equation}{0}
\renewcommand{\theequation}{S\arabic{equation}}
\renewcommand{\thefigure}{S\arabic{figure}}
\newpage 
\clearpage

\section*{SUPPLEMENTARY MATERIAL}

\subsection{Chain representation}
Instead of simulating the dynamics due to the discretized version of Eq.(1-3), we map the bath to a long linear chain with only nearest neighbor coupling with what is known as the ``star-to-chain'' mapping \cite{Wilson1975, prior2010, chin2010b, devega2015b}. 

The Hamiltonian then becomes 
\begin{align}\label{eq:chainunitaryham}
\Hdtp =& \Hop_S + \sum_{j=1}^{N^{\prime}_{\rm max}} \Omega_j\cdop_j \cop_j + \beta_0(\afdop_L \cop_1 + \afop_L\cdop_1) \\ \nonumber
&+ \sum_{j=1}^{N^{\prime}_{\rm max}-1}\beta_j (\cdop_j\cop_{j+1}+\cdop_{j+1}\cop_j),
\end{align}
where $\beta_0 = J\sqrt{\sum_{j=1}^{N_{\rm max}}\spec_j}$. In order to ensure that $\Hop_{\rm tot}^{\rm dis, \prime}$ dictates the same dynamics as $\Hop_{\rm tot}^{\rm dis}$, we used different $\Nm^{\prime}<\Nm$ until the observables converged (we have used $\Nm^{\prime}=200$ except for the case of $N=L=2$ for which we have used $\Nm^{\prime}=400$). To evaluate the other coefficients $\alpha_j$ and $\beta_j$ we have performed a Lanczos tridiagonalization which, via a unitary matrix $U$ converts a diagonal matrix $\Mm=\diag(\omega_1, \omega_2, \dots, \omega_{N_{\rm max}})$ to a tridiagonal matrix $\Tm$ via $\Mm_E \Um \approx \Um \Tm$. The coefficients of $\Tm$ are the $\Omega_j$ and $\beta_j$:  
\begin{align}\label{Tm}
\Tm = \left(\begin{array}{cccc}
\Omega_1 & \beta_1 & 0 & \dots \\
\beta_1 & \Omega_2 & \beta_2 & \dots \\
\vdots & \vdots & \vdots & \vdots \\
0 & 0 & \beta_{N_{\rm max}^{\prime}-1} & \Omega_{N_{\rm max}^{\prime}}
\end{array}\right).
\end{align}
For more details see \cite{devega2015b}. Since the transformation is unitary, the operators $\cop_j$ obey the same bosonic commutation relations as the $\bop_j$.

\subsection{Thermofield transformation and chain representation for environments at finite temperatures}
With the thermofield transformation \cite{devegabanuls2015}, the environment oscillators $\bop_j$ are mapped to $2N_{\rm max}$ oscillators $\aop_{1, j}$ and $\aop_{2, j}$, with the new Hamiltonian $\Hop_{\rm tot}^{{\rm tf}}$
\begin{align}
\Hop_{\rm tot}^{{\rm tf}} =& \Hop_S + \sum_{j=1}^{N_{\rm max}}\omega_j(\adop_{1, j}\aop_{1, j} - \adop_{2, j}\aop_{2, j}) \\ \nonumber
&+\sum_{j=1}^{N_{\rm max}} g_{1, j}(\afdop_L \aop_{1, j} + \afop_L\adop_{1, j}) \\ \nonumber
&+ \sum_{j=1}^{N_{\rm max}} g_{2, j}(\afop_L \aop_{2, j} + \afdop_L\adop_{2, j}),
\end{align}
where $g_{1,j}=J\sqrt{\spec_j}\cosh(\theta_j)$ and $g_{2,j}=J\sqrt{\spec_j}\sinh(\theta_j)$, with $\cosh(\theta_j) = \sqrt{1+n(\omega_j)}$, $\sinh(\theta_j)=\sqrt{n(\omega_j)}$ and $n(\omega)=1/(e^{\;\omega/T}-1)$.
After star-to-chain mapping the Hamiltonian we study is       
\begin{align} \label{eq:H_tf_sc}
\Hop_{\rm tot}^{\rm tf, \prime} =& \Hop_S + \sum_{j=1}^{N^{\prime}_{\rm max}} \Omega_{1,j}\adop_{1,j} \aop_{1,j} + \beta_{1,0}(\afdop_L \aop_{1,1} + \afop_L\adop_{1,1}) \nonumber \\ \nonumber
&+ \sum_{j=1}^{N^{\prime}_{\rm max}-1}\beta_{1,j} (\adop_{1,j}\aop_{1,j+1}+\adop_{1,j+1}\aop_{1,j}) \\ \nonumber 
&+\sum_{j=1}^{N^{\prime}_{\rm max}} \Omega_{2,j}\adop_{2,j} \aop_{2,j} + \beta_{2,0}(\afop_L \aop_{2,1} + \afdop_L\adop_{2,1}) \\ 
&+ \sum_{j=1}^{N^{\prime}_{\rm max}-1}\beta_{2,j} (\adop_{2,j}\aop_{2,j+1}+\adop_{2,j+1}\aop_{2,j}).
\end{align}

\subsection{Redfield master equation}  
We use the following master equation 
\begin{align}\label{eq:nmme2order}
\frac{d\rhoop_S (t)}{dt} =& -\im [H_S, \rhoop_S] + \! \int_0^t d\tau \; \chi^+(\tau)[V_{-\tau} (\afdop_L)\; \rhoop_S(t), \afop_L]  \nonumber  \\  &+ \int_0^t d\tau \chi^-(\tau)[V_{-\tau}(\afop_L)\; \rhoop_S(t), \afdop_L] + \hc ,
\end{align}
where $\chi^-(t) = \sum_{j} \spec_j \left[n(\omega_j) + 1\right]e^{-\im \omega_j t}$, $\chi^+(t) = \sum_{j} \spec_j \; n(\omega_j)e^{\im \omega_j t}$ and the $\omega_j$ are the system eigenvalues.  
We have also used $V_{\tau}(X) = \hat{U}_S^{-1}(\tau, 0)X \hat{U}_S(\tau, 0)$, where $\hat{U}_S(\tau, 0)= e^{-\im \Hop_S \tau}$ \cite{devega2015c}.

\subsection{Two sites and two bosons case}  
In this case only a few energy levels are relevant for the dynamics. As initial condition we take the groundstate for $N=L=2$, whose energy is $E^0_{N=2} = (U-\sqrt{16+U^2})/2$. For $L=2$ and $N=1$ there are only two eigenstates $\vert E^{0, 1}_{N=1}\rangle$ with energies, $E^{0, 1}_{N=1} = \mp 1$, hence only $\vert E^{0}_{N=1}\rangle$ can have energy lower than $E^0_{N=2}$, but only for $U>3$ (because of finite size effects the critical value is $3$ and not $\approx 2.82$). The dynamics between these two states is described by the effective Hamiltonian 
\begin{align}\label{eq:Heff}
\Hop =& \sum_{\omega}  \gamma \sqrt{\spec(\omega)} \left(\vert E^0_{N=2} \rangle \vert 0 \rangle_b \langle E^0_{N=1} \vert \langle \omega \vert_b + \hc \right) \\ \nonumber 
&+ E^0_{N=2} \vert E^0_{N=2}\rangle \langle E^0_{N=2}\vert + E^0_{N=1} \vert E^0_{N=1} \rangle \langle E^0_{N=1} \vert \\ \nonumber 
&+ \sum_{\omega}  \omega \vert \omega \rangle_b \langle \omega \vert_b. 
\end{align}
Here $\vert \omega \rangle_b$ is a state with a single oscillator mode $\omega$ occupied, while $\gamma$ is the coupling between the two states $\vert E^0_{N=2}\rangle$ and $\vert E^0_{N=1}\rangle$ due to $\afdop_2$ acting on the second site. This can be computed by transforming $\afdop_2$ into the eigenbasis of these two number sectors giving $\gamma = \left(4-U+\sqrt{16+U^2}\right)/\left(2\sqrt{16 + U^2 -U\sqrt{16+U^2}}\right)$. Eq.(\ref{eq:Heff}) describes a two-level system coupled with vacuum, which can be solved analytically \cite{devega2015c}.


\begin{thebibliography}{10}
\bibitem{num_cons} In this case, since the total quantum number of the system plus environment is conserved, we use a number conserving MPS time evolution algorithm which respects this symmetry to obtain the best efficiency.
\bibitem{SmallMod} Due to the finite coupling between the system and the bath, especially for small $\Delta E$, the system evolution will not be completely frozen, however, no decay even over long times is expected.
\bibitem{meanfield} Within the Gutzwiller approach, the wavefunction of the system plus bath is described by $\vert \psi\rangle = \otimes_l \left(\sum_n f_{n,l}\vert n\rangle_l\right)$ where $l$ spans all the sites of the system and the bath, while $\vert n\rangle_l$ means $n$ bosons at site $l$. In the Mott-insulating regime, at unit filling the $f_{n,j}$ in the system are given by $f_{n,j}=\delta_{n,1}$ where $\delta_{n,m}$ is the Kronecker $\delta$, while for a $T=0$ bath $f_{n,j}=\delta_{n,0}$. It follows that $\langle \afop_j \rangle = 0 $ and $ \langle \cop_j \rangle = 0$ and hence the dynamics steered by the tunnelling term in the Hamiltonian is completely frozen.
\end{thebibliography}
\end{document}